\documentclass[aps,prd,superscriptaddress,long,notitlepage,balancelastpage,nofootinbib,floatfix,twocolumn]{revtex4-1}
\pdfoutput=1
\usepackage{amsmath,mathtools,amssymb,amsthm,amsxtra,overpic,bbm,epsfig,subfigure,url,bm}
\usepackage{hyperref}
\usepackage{mathrsfs}
\usepackage{color,xcolor}
\usepackage{comment}
\usepackage{float}
\usepackage{enumitem}
\usepackage{orcidlink}
\usepackage[normalem]{ulem}
\usepackage{slashed}
\usepackage{multirow}
\usepackage[normalem]{ulem}
\definecolor{nicered}{rgb}{0.5,0.,0.}
\definecolor{nicegreen}{rgb}{0.,0.5,0.}
\definecolor{niceblue}{rgb}{0.,0.,0.5}
\hypersetup{
	colorlinks=true,
	linkcolor=black,
	filecolor=nicegreen,      
	urlcolor=niceblue,
	citecolor=nicered,
}

\makeatletter
\newcommand*{\balancecolsandclearpage}{%
	\close@column@grid
	\cleardoublepage
	\twocolumngrid
}
\makeatother
\begin{document}

\title{\vspace{1cm} \Large 
Supernova Gamma-Ray Echo as an Astrophysical Near Detector  \\ for Galactic Neutrino Propagation 
}

\author{\bf Garv Chauhan\,\orcidlink{0000-0002-8129-8034}}
\email[E-mail:]{gchauha5@asu.edu}
\affiliation{Department of Physics, Arizona State University, 450 E. Tyler Mall, Tempe, AZ 85287-1504, USA}

\author{\bf Yago Porto\,\orcidlink{0000-0003-3278-0948}}
\email[E-mail:]{yago.porto@tum.de}
\affiliation{Physik-Department, Technische Universit{\"a}t M{\"u}nchen, James-Franck-Stra{\ss}e 1, 85748 Garching, Germany}

\begin{abstract}
Core-collapse supernova neutrinos offer a unique probe of physics beyond the Standard Model over Galactic baselines. Yet even a high-statistics nearby burst may be limited by source uncertainties, especially for propagation effects that average over energy and appear only as an overall flux suppression. Inspired by near--far comparisons in terrestrial oscillation experiments, we show that next-generation MeV observations of the supernova gamma-ray echo---the delayed 511 keV signal induced by the neutrino burst in the stellar envelope---can serve as an astrophysical near detector for the emitted $\bar\nu_e$ fluence. Comparing this echo-inferred source fluence with the surviving flux measured at Earth gives a direct test of neutrino propagation between the stellar surface and the detector. For pseudo-Dirac neutrinos, this turns the fully averaged active--sterile oscillation regime from an unobservable source-normalization ambiguity into a measurable near--far deficit, with sensitivity for nearby progenitors to splittings
$\delta m^2\gtrsim 10^{-18}~{\rm eV}^2(0.22~{\rm kpc}/D)(E_\star/15~{\rm MeV})$.
The same near--far logic also applies to other propagation-induced modifications of the observed flux, including neutrino decay and neutrino interactions with Galactic dark matter.

\noindent 
\end{abstract}
\maketitle
\textbf{\emph{Introduction}.--} 
The next Galactic core-collapse supernova will be a major laboratory for neutrino physics~\cite{Mirizzi:2015eza}. Its neutrino burst is expected to yield very large event samples in current and next-generation detectors, enabling detailed studies of both supernova dynamics and neutrino properties~\cite{Super-Kamiokande:2002weg, IceCube:2011cwc, JUNO:2015zny, Hyper-Kamiokande:2018ofw, DUNE:2020zfm}. For a sufficiently nearby source, the combination of enormous statistics and an astrophysical baseline would also provide sensitivity to extremely feeble new-physics effects accumulated during propagation.

Yet sensitivity to such new physics is not guaranteed by high event statistics alone. Many long-baseline effects leave no sharp spectral distortion or distinctive time-dependent signature, but instead appear as an approximately energy-smooth suppression of the flux observed at Earth. Without an independent probe of the emitted fluence, such a deficit is degenerate with the uncertain supernova emission normalization.

In terrestrial oscillation experiments, analogous normalization uncertainties are controlled by measuring the unoscillated flux near the source and comparing it with the flux observed after propagation~\cite{DayaBay:2012fng}. For sufficiently nearby supernovae, next-generation MeV observations of the gamma-ray echo provides the corresponding near measurement.
This echo is a 511 keV signal produced when the emitted $\bar\nu_e$ burst undergoes inverse beta decay in the stellar envelope, followed by positron annihilation~\cite{Bisnovatyi-Kogan1975, Ryazhskaya1999, Lu:2007wp, Lunardini:2023ilg,Chauhan:2026nqf}.
It therefore probes the $\bar\nu_e$ fluence emerging from the stellar surface, before any additional propagation effect between the star and Earth, while terrestrial neutrino detectors measure the propagated burst. The comparison defines an astrophysical near--far test of long-baseline neutrino propagation.

\textbf{\emph{Gamma-ray echo as a near detector}.--} 
The gamma-ray echo is the delayed 511~keV transient signal produced when the outgoing supernova $\bar\nu_e$ burst crosses the hydrogen-rich outer envelope of the progenitor. Inverse beta decay, $\bar\nu_e+p\to n+e^+$, followed by rapid positron thermalization and annihilation at rest, converts part of the $\bar\nu_e$ fluence into 511~keV photons emitted near the stellar surface. These photons arrive at Earth with geometric delays across the visible stellar disk, producing a transient signal whose duration is set by the stellar light-crossing time, $\Delta t\sim R_\star/c$~\cite{Lunardini:2023ilg}. Since the echo is generated in the stellar envelope, it probes the $\bar\nu_e$ fluence after flavor conversion inside the supernova but before any additional propagation effect between the stellar surface and Earth.

The key property that makes the echo a useful near detector is that its photon yield is directly proportional to the emitted $\bar\nu_e$ fluence, with a coefficient controlled by the physics of the outer envelope. Following Ref.~\cite{Lunardini:2023ilg}, we approximate the contributing region as the surface layer from which 511~keV photons can escape without Compton scattering, with corresponsing cross-section denoted by $\sigma_C$. A photon produced at radius $r$ lies within this layer when the radial optical depth $\tau_C(r)$ is smaller than or of order unity. The corresponding electron column is therefore $\simeq{1}/{\sigma_C}$. Writing $Y_p$ and $Y_e$ for the free-proton and electron number fractions per baryon, respectively, and taking $Y_p/Y_e$ to be approximately uniform across this layer, its corresponding proton column is $\simeq{Y_p}/({Y_e}\,{\sigma_C})$. A radially outgoing $\bar\nu_e$ crossing this layer consequently undergoes inverse beta decay with probability $P_{\rm IBD}(E_\nu)\simeq({Y_p}\,{\sigma_{\rm IBD}})/({Y_e}\,{\sigma_C})$. The resulting positron rapidly thermalizes and annihilates, producing on average $\eta_\gamma\simeq1.74$ photons in the 511~keV line~\cite{Lu:2007wp}. In the surface-layer approximation, the inward-going photons are absorbed, while the outward-going half are taken to escape. The number of escaping line photons produced per incident $\bar\nu_e$ is therefore $\simeq({\eta_\gamma\,Y_p}\,{\sigma_{\rm IBD}})/(2\,{Y_e}\,{\sigma_C})$. The factor $1/2$ accounts for the fact that isotropic emission sends, on average, half of the photons outward and half inward.

Integrating over the full $\bar\nu_e$ emission and accounting for geometric dilution and the telescope effective area $A_{\rm eff}$, the total number of extracted 511~keV echo counts is
\begin{equation}
N_\gamma^\oplus
=
\frac{A_{\rm eff}}{4\pi D^2}\,
C_\gamma
\int dE_\nu\,dt\;
\sigma_{\rm IBD}(E_\nu)\,
\Phi_{\bar\nu_e}^{\rm surf}(E_\nu,t),
\label{eq:Ngamma}
\end{equation}
where $D$ is the distance from Earth to the progenitor, $\Phi_{\bar\nu_e}^{\rm surf}(E_\nu,t)\equiv d^2N_{\bar\nu_e}^{\rm surf}/dE_\nu\,dt$ is the post-flavor-conversion $\bar\nu_e$ differential number luminosity at the stellar surface, and
\begin{equation}
C_\gamma
\equiv
\frac{\eta_\gamma}{2}\,
\frac{Y_p}{Y_e}\,
\frac{1}{\sigma_C}
\label{eq:Cgamma}
\end{equation}
collects the envelope and photon-escape factors.\footnote{\label{fn:photon_transport}The normalization of Ref.~\cite{Lu:2007wp} corresponds to an effective photon-escape factor $1/8$, four times smaller than the $1/2$ surface-layer value used here. This spread illustrates the present photon-transport modeling uncertainty in $C_\gamma$.} Eq.~\eqref{eq:Ngamma} makes the near-detector role of the echo explicit: $N_\gamma^\oplus$ is directly proportional to the source $\bar\nu_e$ fluence, up to the calibration factor $C_\gamma$. In writing Eq.~\eqref{eq:Ngamma}, we assume spherically symmetric $\bar\nu_e$ emission, so that the fluence crossing the Earth-facing side of the envelope is the same as elsewhere on the stellar surface. The consequences of large-scale emission anisotropy for the echo-based near--far comparison are studied in Ref.~\cite{Chauhan:2026nqf}. The dominant uncertainty in using $N_\gamma^\oplus$ as a fluence proxy is therefore the calibration of $C_\gamma$, set by the envelope composition and photon-production and escape physics.

\textbf{\textit{Far-detector signal}.}--
The corresponding far measurement is the $\bar\nu_e$ burst observed in a terrestrial IBD detector after propagation from the stellar surface. Writing the event yield in the same form as Eq.~\eqref{eq:Ngamma},
\begin{equation}
N_{\rm IBD}^{\oplus}
=
\frac{N_p\,\varepsilon_{\rm IBD}}{4\pi D^2}
\int dE_\nu\,dt\;
\sigma_{\rm IBD}(E_\nu)\;
S(E_\nu)\;
\Phi_{\bar\nu_e}^{\rm surf}(E_\nu,t),
\label{eq:Nibd}
\end{equation}
where $N_p$ is the number of free proton targets, $\varepsilon_{\rm IBD}$ is an effective IBD detection efficiency, taken here to be approximately energy independent, and $S(E_\nu)$ parametrizes any additional propagation effect that changes the $\bar\nu_e$ flux between the stellar surface and Earth. In the standard three-flavor framework, after ordinary flavor conversion inside the supernova has already been absorbed into $\Phi_{\bar\nu_e}^{\rm surf}$, one has $S=1$. Values $S<1$ describe attenuation or conversion away from the active $\bar\nu_e$ flux, while $S>1$ would correspond to enhancement of the detected $\bar\nu_e$ component during propagation. Comparing Eqs.~\eqref{eq:Ngamma} and~\eqref{eq:Nibd} makes the near--far structure transparent: both observables are driven by the same post-flavor-conversion surface emission and the same IBD cross section, while only the terrestrial signal carries the propagation factor $S(E_\nu)$. The echo therefore supplies the source-fluence normalization needed to test long-baseline propagation.

\textbf{\emph{Pseudo-Dirac benchmark}.--}
Among the scenarios that benefit most directly from the near--far echo comparison are pseudo-Dirac neutrinos, in which each active state is paired with a nearly degenerate sterile partner and mixes maximally with it, $\theta_s \approx \pi/4$, producing oscillations with maximal amplitude and extremely long wavelengths set by tiny mass-squared splittings~\cite{Wolfenstein:1981kw, deGouvea:2009fp}. Assuming, for simplicity, a common active--sterile splitting $\delta m^2$ for the three pairs, and after the standard oscillation phases have averaged out, the additional propagation factor in Eq.~\eqref{eq:Nibd} is
\begin{equation}
S(E_\nu)
=
\cos^2\!\left(\frac{\delta m^2 D}{4E_\nu}\right),
\label{eq:pseudo_dirac_S}
\end{equation}
where $D$ is the supernova distance. Once the pseudo-Dirac phase is itself averaged, $S(E_\nu)\to1/2$. The terrestrial neutrino signal is then reduced by an energy-independent factor of two, producing a shape-preserving normalization deficit that is degenerate with the uncertain supernova emission normalization in the neutrino signal alone. The gamma-ray echo provides the independent source-fluence measurement needed to turn this averaged suppression into a near--far observable.

\begin{figure}[!t]
    \centering
    \includegraphics[width=0.95\linewidth]{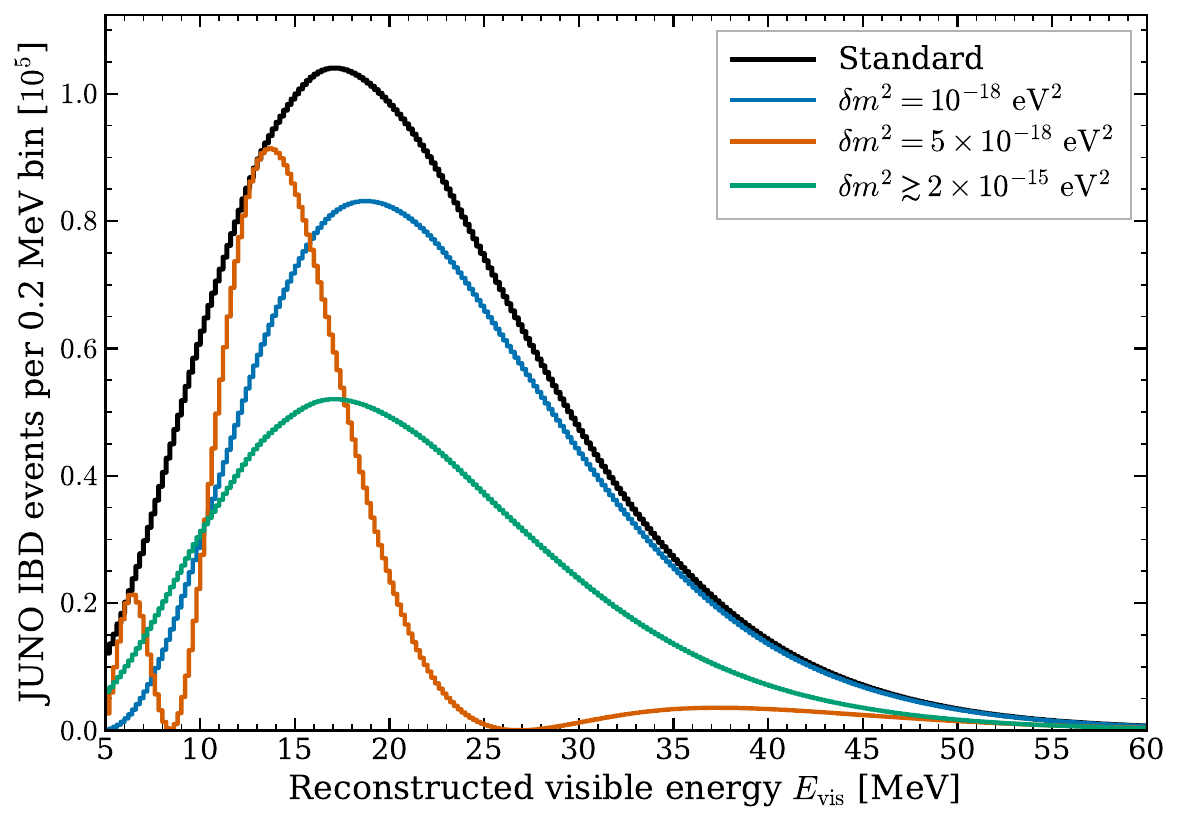}
    \caption{
    Reconstructed visible-energy spectrum of selected inverse beta decay events in JUNO for a Galactic supernova at $D=0.222~\mathrm{kpc}$, assuming complete flavor equipartition at the stellar surface.
    The normalization uses $N_p=1.44\times10^{33}$ and $\varepsilon_{\rm IBD}=0.822$.
    The black solid histogram shows the standard three-neutrino prediction, while the blue and orange histograms show pseudo-Dirac scenarios with
    $\delta m^2=10^{-18}~\mathrm{eV}^2$ and
    $\delta m^2=5\times10^{-18}~\mathrm{eV}^2$, respectively.
    The green solid histogram shows the fully averaged pseudo-Dirac limit, $S=1/2$, reached to within a few percent for
    $\delta m^2\gtrsim2\times10^{-15}~\mathrm{eV}^2$ in this Betelgeuse/JUNO setup.
    The spectra are shown in 0.2~MeV bins as a function of reconstructed visible energy $E_{\rm vis}$.
    }
    \label{fig:juno_pd_spectrum}
\end{figure}

The gamma-ray echo is detectable only for nearby progenitors, roughly within $D\lesssim 1~{\rm kpc}$ for next-generation MeV telescopes~\cite{Lunardini:2023ilg,Chauhan:2026nqf}. This range includes candidates such as Betelgeuse ($D\simeq0.22~{\rm kpc}$)~\cite{Harper_2017} and Rigel ($D\simeq0.26~{\rm kpc}$)~\cite{10.1093/mnras/stac1617}. Setting the pseudo-Dirac phase in Eq.~\eqref{eq:pseudo_dirac_S} to order unity across the relevant burst-energy range gives the characteristic splitting
\begin{equation}
\delta m^2
\sim
10^{-18}~{\rm eV}^2
\left(\frac{0.22~{\rm kpc}}{D}\right)
\left(\frac{E_\star}{15~{\rm MeV}}\right).
\label{eq:pd_scale}
\end{equation}
For splittings above this scale, pseudo-Dirac oscillations develop across the burst spectrum and eventually average to a normalization deficit. Fig.~\ref{fig:juno_pd_spectrum} illustrates this transition for a Betelgeuse-like burst in the Jiangmen Underground Neutrino Observatory (JUNO)~\cite{JUNO:2015zny}. The calculation uses the Garching SFHo supernova-neutrino model with final baryonic neutron-star mass $1.93\,M_\odot$~\cite{GarchingArchive,Fiorillo:2023frv} and the IBD cross section implemented in SNOwGLoBES~\cite{SNOwGLoBES}. We model the JUNO energy response with a Gaussian resolution, $\sigma_E/E_{\rm vis}=3\%/\sqrt{E_{\rm vis}/{\rm MeV}}$~\cite{JUNO:2021vlw}. This excellent energy resolution makes JUNO particularly well suited to resolving the oscillatory spectral features produced by pseudo-Dirac propagation. For the normalization, we use $N_p=1.44\times10^{33}$ free protons and $\varepsilon_{\rm IBD}=0.822$~\cite{JUNO:2024jaw}, consistent with the $0.81$--$0.85$ selection fractions found in JUNO supernova-monitor simulations~\cite{JUNO:2023dnp}. For a Betelgeuse-distance burst, this efficiency should be regarded as a benchmark, since the high event rate and pile-up require dedicated detector simulation and reconstruction~\cite{JUNO:2023dnp}. For the $\bar\nu_e$ fluence at the stellar surface, we adopt complete flavor equipartition, $\Phi_{\bar\nu_e}^{\rm surf}=(\Phi_{\bar\nu_e}^{0}+2\Phi_{\bar\nu_x}^{0})/3$, where $\Phi_{\bar\nu_\alpha}^{0}$ denotes the unoscillated source fluence and $\Phi_{\bar\nu_x}^{0}$ denotes the fluence per heavy-lepton antineutrino flavor, with $x=\mu,\tau$. This provides a well-defined benchmark, while the detailed flavor transformation inside the supernova remains model dependent. The near--far observable is independent of this choice, since any flavor conversion inside the star is already absorbed into $\Phi_{\bar\nu_e}^{\rm surf}$.

In Fig.~\ref{fig:juno_pd_spectrum}, the standard spectrum is compared with two finite pseudo-Dirac splittings and the fully averaged limit, chosen to show the progression from a weakly distorted spectrum to a resolved oscillatory feature and finally to the practically averaged regime. For $\delta m^2=10^{-18}~{\rm eV}^2$, the oscillation phase varies too slowly across the relevant energy range to produce a distinctive spectral feature, leaving only a mild, smooth suppression and making a diagnosis from the neutrino spectrum alone improbable. For $\delta m^2=5\times10^{-18}~{\rm eV}^2$, the oscillation minimum falls in the high-statistics part of the spectrum, producing a large distortion that represents a best-case scenario for spectral identification. At larger splittings, the oscillations become increasingly rapid and are progressively washed out by the detector response and 0.2~MeV reconstructed-energy binning. For $\delta m^2\gtrsim2\times10^{-15}~{\rm eV}^2$, the reconstructed spectrum agrees with the fully averaged prediction, $S=1/2$, at the few-percent level. Worse energy resolution or coarser binning would move the onset of practical averaging to smaller splittings, shrinking the window where spectral identification alone is viable and making the near--far comparison correspondingly more important.

\textbf{\textit{Near--far estimator and reach}.}--
Comparing Eqs.~\eqref{eq:Ngamma} and~\eqref{eq:Nibd} motivates the dimensionless near--far estimator
\begin{equation}
\kappa
\equiv
\frac{N_{\rm IBD}^{\oplus}}{N_{\gamma}^{\oplus}}\,
\frac{A_{\rm eff}C_\gamma}{N_p\,\varepsilon_{\rm IBD}}.
\label{eq:kappa}
\end{equation}
Because both $N_{\rm IBD}^\oplus$ and $N_\gamma^\oplus$ scale as $D^{-2}$, $\kappa$ is independent of the source distance. In the standard scenario with $S(E_\nu)=1$, one expects $\kappa=1$ up to experimental uncertainties. If the propagation suppression is approximately energy independent over the IBD-weighted burst spectrum, $S(E_\nu)\simeq S$, then $\kappa\simeq S$. This is precisely the situation realized by fully averaged pseudo-Dirac oscillations, for which $S\simeq1/2$. Otherwise, $\kappa$ measures the corresponding IBD-weighted average suppression.

To quantify the sensitivity, we propagate the dominant uncertainties in $\kappa$. The far-detector count has Poisson uncertainty $\sigma_{N_{\rm IBD}}\simeq\sqrt{N_{\rm IBD}^\oplus}$. The echo measurement requires extracting a signal $N_\gamma^\oplus$ above the diffuse Galactic 511~keV line background $N_b$ within the selected sky region, giving $\sigma_{N_\gamma}\simeq\sqrt{N_\gamma^\oplus+N_b}$. Combining these counting terms with uncorrelated systematic uncertainties, and defining $\beta\equiv N_b/N_\gamma^\oplus$, gives
\begin{equation}
\left(\frac{\sigma_\kappa}{\kappa}\right)^2
\simeq
\frac{1}{N_{\rm IBD}^{\oplus}}
+
\frac{1+\beta}{N_{\gamma}^{\oplus}}
+
\sigma_{\rm sys}^2,
\label{eq:kappa_unc}
\end{equation}
where
\begin{equation}
\sigma_{\rm sys}^2
\equiv
\left(\frac{\sigma_{A_{\rm eff}}}{A_{\rm eff}}\right)^2
+
\left(\frac{\sigma_{C_\gamma}}{C_\gamma}\right)^2
+
\left(\frac{\sigma_{N_p}}{N_p}\right)^2
+
\left(\frac{\sigma_{\varepsilon_{\rm IBD}}}{\varepsilon_{\rm IBD}}\right)^2.
\label{eq:sigmasys}
\end{equation}
An $n\sigma$ departure from $\kappa=1$ corresponds to
\begin{equation}
|\kappa-1|
\gtrsim
n\,\sigma_\kappa
=
n\,\kappa
\sqrt{
\frac{1}{N_{\rm IBD}^{\oplus}}
+
\frac{1+\beta}{N_{\gamma}^{\oplus}}
+
\sigma_{\rm sys}^2
}.
\label{eq:departure_condition}
\end{equation}
For the pseudo-Dirac case of interest here, the effect is a deficit, $\kappa<1$. For the nearby supernovae relevant to an observable echo, the terrestrial IBD sample is enormous ($N_{\rm IBD}^{\oplus}\simeq1.37\times10^7$ and $1/N_{\rm IBD}^{\oplus}\simeq7\times10^{-8}$ for our benchmark), making its counting contribution negligible compared with the echo counting term and systematic uncertainties. Dropping this term, Eq.~\eqref{eq:departure_condition} gives
\begin{equation}
\kappa
\lesssim
\left[
1+n
\sqrt{
\frac{1+\beta}{N_{\gamma}^{\oplus}}
+
\sigma_{\rm sys}^2
}
\right]^{-1}.
\label{eq:kappalimit}
\end{equation}
Eq.~\eqref{eq:kappalimit} gives the largest value of $\kappa<1$ that can be resolved at $n\sigma$, with the threshold moving closer to unity as the echo statistics improve or $\sigma_{\rm sys}$ decreases.

\textbf{\textit{Resolving the averaged pseudo-Dirac deficit}.}--
The natural target for the near--far comparison is the fully averaged pseudo-Dirac regime, for which $\kappa\simeq S\simeq1/2$. Setting $\kappa=1/2$ in Eq.~\eqref{eq:kappalimit} gives, for an $n\sigma$ distinction from the standard expectation,
\begin{equation}
\frac{1+\beta}{N_\gamma^\oplus}
+
\left(\frac{\sigma_{A_{\rm eff}}}{A_{\rm eff}}\right)^2
+
\left(\frac{\sigma_{C_\gamma}}{C_\gamma}\right)^2
\lesssim
\frac{1}{n^2}.
\label{eq:half_deficit_requirement}
\end{equation}
The first term is the counting variance of the background-subtracted echo measurement, while the remaining terms describe the effective-area and echo-normalization uncertainties. JUNO estimates the combined normalization uncertainty from $N_p$ and $\varepsilon_{\rm IBD}$ at approximately $1\%$, dominated by the $0.9\%$ target-proton contribution~\cite{JUNO:2024jaw}. These uncertainties are subdominant, so we neglect the corresponding terms in Eq.~\eqref{eq:sigmasys}.

The echo statistics are determined directly from Eq.~\eqref{eq:Ngamma}, using the same complete-flavor-equipartition benchmark and IBD cross section employed in Fig.~\ref{fig:juno_pd_spectrum}. We take $D=0.222~{\rm kpc}$ and adopt as a representative next-generation soft-MeV instrument an AMEGO--like effective area $A_{\rm eff}=3\times10^3~{\rm cm}^2$~\cite{AMEGO:2019gny,Kierans:2020otl,AMEGOTechnical}. The surface-layer prescription in Eq.~\eqref{eq:Cgamma} gives $C_\gamma=2.49\times10^{24}~{\rm cm}^{-2}$, using $\eta_\gamma=1.74$~\cite{Lu:2007wp}, the Klein--Nishina cross section $\sigma_C=2.87\times10^{-25}~{\rm cm}^2$ at 511~keV~\cite{doi:https://doi.org/10.1002/9783527618170.ch7}, and $Y_p/Y_e\simeq0.822$ for the hydrogen-rich Betelgeuse surface composition~\cite{Dolan_2016}. For the $20\,M_\odot$ benchmark, which lies within the inferred Betelgeuse initial-mass range of $18$--$21\,M_\odot$~\cite{Joyce_2020}, Eq.~\eqref{eq:Ngamma} gives $N_\gamma^\oplus=70.7$. Applying the same calculation to the $9\,M_\odot$ Garching model gives $N_\gamma^\oplus=29.7$, illustrating the dependence of the echo statistics on the progenitor model. 

The Galactic background accumulated over the echo duration is estimated as $N_b=\Phi_{\rm gal}A_{\rm eff}R_\star/c$, where $\Phi_{\rm gal}$ is the diffuse Galactic 511~keV line flux within the sky region used to extract the echo. Integrating the two-dimensional Galactic-disk model of Ref.~\cite{Siegert:2015knp} over a $3^\circ$-radius region centered on Betelgeuse gives $\Phi_{\rm gal}=2.4\times10^{-7}~{\rm cm}^{-2}\,{\rm s}^{-1}$. Using $R_\star=764^{+116}_{-62}\,R_\odot$~\cite{Joyce_2020}, we obtain $N_b=1.28$, with the radius uncertainty spanning $N_b=1.17$--$1.47$. The corresponding background-to-signal ratios are $\beta=0.0181$ for the $20\,M_\odot$ model and $\beta=0.0429$ for the $9\,M_\odot$ model. This estimate includes only the Galactic line background, since continuum and instrumental contributions depend on the detector and adopted energy window.

We adopt $\sigma_{A_{\rm eff}}/A_{\rm eff}=0.05$ as a representative benchmark for a future soft-MeV telescope.\footnote{Laboratory calibration of the soft-MeV Compton telescope COSI found a $1.6\%$ flux-normalization systematic for the 511~keV line~\cite{Sleator:2019dks}. We adopt $5\%$ to allow for additional calibration uncertainty under realistic observing conditions.} For the $20\,M_\odot$ model, the counting and effective-area terms are $(1+\beta)/N_\gamma^\oplus\simeq0.0144$ and $(\sigma_{A_{\rm eff}}/A_{\rm eff})^2=0.0025$, respectively.
The resulting calibration requirements are
\begin{equation}
\frac{\sigma_{C_\gamma}}{C_\gamma}
\lesssim
48.3\%\quad(2\sigma),
\qquad
\frac{\sigma_{C_\gamma}}{C_\gamma}
\lesssim
30.7\%\quad(3\sigma).
\label{eq:Cgamma_target}
\end{equation}
At $1\sigma$, the corresponding calibration requirement is $\sigma_{C_\gamma}/C_\gamma\lesssim99.2\%$. Nevertheless, the factor-of-four spread between the photon-escape prescriptions discussed in footnote~\ref{fn:photon_transport} is broad enough to absorb the factor-of-two pseudo-Dirac deficit itself, preventing a robust interpretation with the current modeling. Because this spread originates from simplified treatments of photon escape, dedicated transport calculations should narrow it substantially and make the $2\sigma$ and $3\sigma$ targets achievable.

\begin{figure*}[t!]
    \centering
    \includegraphics[width=0.65\textwidth]{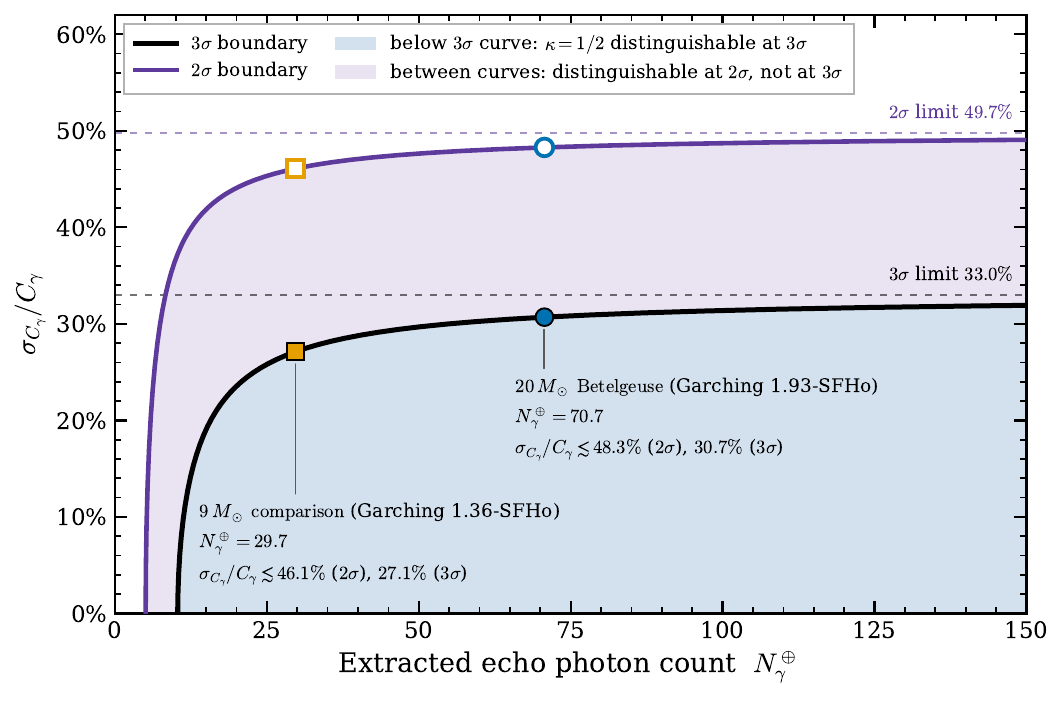}
    \caption{
Requirement for resolving the fully averaged pseudo-Dirac deficit, $\kappa=1/2$, in the near--far comparison.
The black and purple curves show the $3\sigma$ and $2\sigma$ boundaries from Eq.~\eqref{eq:half_deficit_requirement}, evaluated for $N_b=1.28$ and $\sigma_{A_{\rm eff}}/A_{\rm eff}=5\%$.
The region below the black curve corresponds to $3\sigma$ sensitivity, while the region between the black and purple curves corresponds to $2\sigma$ sensitivity.
The blue circles and orange squares mark the $20\,M_\odot$ Betelgeuse and $9\,M_\odot$ comparison benchmarks, respectively.
The dashed horizontal lines show the corresponding large-statistics limits.
}
    \label{fig:kappa_requirement}
\end{figure*}

Fig.~\ref{fig:kappa_requirement} summarizes the joint requirement on echo statistics and $C_\gamma$ calibration. The lower shaded region marks where $\kappa=1/2$ can be distinguished from $\kappa=1$ at $3\sigma$, The black and purple curves delimit the regions where the deficit is resolvable at $3\sigma$ and $2\sigma$, respectively. The boundaries are evaluated for $N_b=1.28$ and $\sigma_{A_{\rm eff}}/A_{\rm eff}=5\%$, with $\beta=N_b/N_\gamma^\oplus$ varying along the curves. For the $20\,M_\odot$ Betelgeuse benchmark, $N_\gamma^\oplus=70.7$ gives maximum uncertainties of $48.3\%$ at $2\sigma$ and $30.7\%$ at $3\sigma$. For the $9\,M_\odot$ model, $N_\gamma^\oplus=29.7$ gives $46.1\%$ and $27.1\%$, respectively. At low photon counts the boundaries are controlled by counting statistics. As $N_\gamma^\oplus$ increases, they approach the large-statistics limits of $49.7\%$ at $2\sigma$ and $33.0\%$ at $3\sigma$. The Betelgeuse benchmark already lies close to these limits, so additional echo photons provide limited improvement and control of the $C_\gamma$ normalization remains the central requirement.

The precision required on $C_\gamma$ in Eq.~\eqref{eq:Cgamma_target} is modest.
Because averaged pseudo-Dirac oscillations halve the detected flux, a
tens-of-percent prediction of $C_\gamma$ suffices even after including echo
counting statistics and the adopted $5\%$ effective-area uncertainty. By
contrast, the companion anisotropy analysis targets near--far deviations of
only $10$--$15\%$, which demand larger effective areas and much tighter
normalization control~\cite{Chauhan:2026nqf}. The principal remaining
theoretical uncertainty is the factor-of-four spread between photon-escape
prescriptions discussed in footnote~\ref{fn:photon_transport}, which reflects
their simplified treatment of photon transport rather than an intrinsic
limitation of the method. The averaged pseudo-Dirac deficit is therefore a
realistic first target for the echo near--far comparison, and dedicated
stellar and transport calculations should sharpen the test further. In the
longer term, the 2.22~MeV line from neutron capture on protons produced by the
same IBD events~\cite{Lu:2007wp} could offer an independent cross-check of
$C_\gamma$, although its fluence is roughly two orders of magnitude lower and
would require a correspondingly larger effective area.


\textbf{\textit{Discussion and outlook}.}--
We have shown that the gamma-ray echo of a nearby supernova can act as an
astrophysical near detector, anchoring the emitted $\bar\nu_e$ fluence and
thereby turning energy-smooth propagation effects into a near--far observable.
Applying the method to an actual event defines a clear program of
target-specific echo and background calculations, dedicated 511~keV
photon-transport modeling of the outer envelope, and a well-characterized
soft-MeV detector response. The tens-of-percent precision identified here
provides a concrete target for these efforts.

Traditional spectral searches for pseudo-Dirac oscillations~\cite{Martinez-Soler:2021unz,Carloni:2022cqz,Carloni:2025dhv} are sensitive only over a finite range of mass splittings, where the oscillation phase is large enough to produce a visible energy-dependent modulation but not yet large enough for that modulation to average out. At larger splittings the oscillations average to a featureless factor-of-two deficit, causing spectral searches to lose sensitivity because the deficit is degenerate with the uncertain source normalization. The echo breaks this degeneracy by anchoring the emitted fluence, allowing the near--far comparison to probe the averaged regime. Combining the spectral and near--far information therefore gives sensitivity to
$\delta m^2\gtrsim10^{-18}~{\rm eV}^2(0.22~{\rm kpc}/D)(E_\star/15~{\rm MeV})$, as summarized in Eq.~\eqref{eq:pd_scale}.

The same normalization test probes other propagation effects that produce an approximately energy-smooth suppression. A factor-of-two deficit corresponds to $\tau/m\lesssim2.2\times10^3~{\rm s/eV}(D/0.22~{\rm kpc})(15~{\rm MeV}/E_\star)$ for invisible neutrino decay and $\sigma_{\nu\chi}/m_\chi\gtrsim2.7\times10^{-21}~{\rm cm^2\,GeV^{-1}}(0.22~{\rm kpc}/D)(0.39~{\rm GeV\,cm^{-3}}/\rho_{\chi,\odot})$ for neutrino--dark-matter scattering. The near--far test complements dedicated spectral, high-energy, and diffuse-flux searches~\cite{Ivanez-Ballesteros:2023lqa,Valera:2024buc,Chauhan:2025hoz,Esteban:2025wbv}, while depending much less on assumptions about the source fluence. More general effects need not produce a deficit, and visible neutrino decay can instead give $\kappa>1$ depending on the mass ordering and decay channel.

\vspace{0.2in}
\textbf{\emph{Acknowledgments}.--}
The work of G.C. is supported by NSF Awards Number 2309973 and 2609687. YP is supported by the Collaborative Research Center SFB1258 and by the Deutsche Forschungsgemeinschaft (DFG, German Research Foundation) under Germany's Excellence Strategy - EXC-2094 - 390783311. G.C. and Y.P. thank the Center for Theoretical Underground Physics and Related Areas (CETUP* 2025 and CETUP* 2026) and the Institute for Underground Science at SURF for hospitality and for providing a stimulating environment, where part of this work was carried out.

\bibliographystyle{apsrev4-1}
\bibliography{ref}
\end{document}